\documentclass[reprint, twocolumn, amsmath,amssymb]{revtex4}
\usepackage{array}
\newcommand{\PreserveBackslash}[1]{\let\temp=\\#1\let\\=\temp}
\newcolumntype{C}[1]{>{\PreserveBackslash\centering}p{#1}}
\newcolumntype{R}[1]{>{\PreserveBackslash\raggedleft}p{#1}}
\newcolumntype{L}[1]{>{\PreserveBackslash\raggedright}p{#1}}

\usepackage{graphicx}% Include figure files
\usepackage{dcolumn}% Align table columns on decimal point
\usepackage{bm}% bold math
\usepackage{xcolor}
\usepackage{siunitx}
\usepackage{booktabs}
\usepackage[hidelinks]{hyperref}
\usepackage{orcidlink}

\newcommand{\mres}{\ensuremath{N \lambda/4} }
\newcommand{\qres}{\ensuremath{\lambda / 4} }

\newcommand{\E}{\ensuremath{\vec{E}}}

\newcommand{\nph}{\ensuremath{n_{\mathrm{ph},\lambda/4~} }}

\makeatletter 
    
\renewcommand\onecolumngrid{% <<<<<<
\do@columngrid{one}{\@ne}%
\def\set@footnotewidth{\onecolumngrid}% <<<<<<<<<<<<<<<<
\def\footnoterule{\kern-6pt\hrule width 1.5in\kern6pt}%
}

\renewcommand\twocolumngrid{% <<<<<<
        \def\footnoterule{% restore rule
        \dimen@\skip\footins\divide\dimen@\thr@@
        \kern-\dimen@\hrule width.5in\kern\dimen@}
        \do@columngrid{mlt}{\tw@}
}%

\makeatother 

\begin{document}

\title{Broadband and high-precision two-level system loss measurement using superconducting multi-wave resonators}

\begin{abstract}
Two-level systems (TLS) are known to be a dominant source of dissipation and decoherence in superconducting qubits. Superconducting resonators provide a convenient way to study TLS-induced loss due to easier design and fabrication in comparison to devices that include non-linear elements such as Josephson junctions. However, accurately measuring TLS-induced loss in a resonator in the quantum regime is challenging due to low signal-to-noise ratio (SNR) and the temporal fluctuations of the TLS, leading to uncertainties of 30\% or more in the loss measurement.  To address these limitations, we develop a multi-wave resonator device that extends the resonator length from a standard quarter-wave $\lambda/4$ to $N\lambda/4$ where $N = 37$ at \SI{6}{\GHz}. This design provides two key advantages: the TLS-induced fluctuations are reduced by a factor of $\sqrt{N}$ due to spatial averaging over an increased number of independent TLS, and the measurement SNR for a given intra-resonator energy density improves by a factor of $\sqrt{N}$ as well. The multi-wave resonator also has fundamental and harmonic resonances spanning a wide frequency range, allowing one to study the frequency dependence of TLS-induced loss. In this work we fabricate both multi-wave and quarter-wave coplanar waveguide resonators formed from thin-film aluminum on a silicon substrate, and characterize their TLS properties at both \SI{10}{\milli\kelvin} and \SI{200}{\milli\kelvin}. Our results show that the power-dependent TLS-induced loss measured from the $N\lambda/4$ resonator agrees well with the $\lambda/4$ resonators, while achieving a five-fold reduction in measurement uncertainty due to TLS fluctuations, down to $5\%$. The $N\lambda/4$ resonator also provides a measure of the fully unsaturated TLS-induced loss due to the improved measurement SNR at low intra-resonator energy densities. Finally, measurements across seven harmonic resonances of the $N\lambda/4$ resonator between \qtyrange[range-phrase = ~--~]{4}{6.5}{\GHz} reveals no frequency dependence in the TLS-induced loss over this range. These results highlight the benefits of the multi-wave resonator approach for materials development in superconducting quantum circuits. 
\end{abstract}
 
\author{Cliff Chen\,\orcidlink{0009-0009-1424-2039}}
\email{clfchen@amazon.com}
\affiliation{AWS Center for Quantum Computing, Pasadena, CA 91106, USA}
\author{Shahriar Aghaeimeibodi\,\orcidlink{0000-0002-9920-493X}}
\affiliation{AWS Center for Quantum Computing, Pasadena, CA 91106, USA}
\author{Yuki Sato}
\affiliation{AWS Center for Quantum Computing, Pasadena, CA 91106, USA}
\author{Matthew H. Matheny\,\orcidlink{0000-0002-3488-1083}}
\affiliation{AWS Center for Quantum Computing, Pasadena, CA 91106, USA}
\author{Oskar Painter\,\orcidlink{0000-0002-1581-9209}}
\affiliation{AWS Center for Quantum Computing, Pasadena, CA 91106, USA}
\author{Jiansong Gao\,\orcidlink{0009-0005-2728-9371}}
\affiliation{AWS Center for Quantum Computing, Pasadena, CA 91106, USA}

\date{\today}
\maketitle

\section{Introduction}
Superconducting qubits are among the most promising platforms for scalable quantum computing due to their compatibility with semiconductor technology \cite{Wang2022, Ezratty2023}. Recent advances in quantum error correction experiments highlight the need for high coherence qubits in order to surpass the threshold for fault-tolerant quantum computation \cite{Putterman2025, google2024}. Coherence times have steadily increased over the years in large part due to improved circuit designs that reduce the influence of two-level system (TLS) defects located at the interfaces between substrate, metal, and air \cite{Muller2019, Kjaergaard2020, Martinis2005, Gao2008_3, Lisenfeld2016, Woods2019}. These defects are generally assumed to couple the qubit's electric field, and recent advances have focused on reducing this coupling either through optimized design of the qubit capacitor or through the choice of superconducting thin-films with lower surface-interface defect density and corresponding loss tangent~\cite{Muller2019, MURRAY2021, Place2021}. 

Superconducting resonators are a popular way to assess this TLS-induced loss due to their simpler fabrication and measurement requirements \cite{Mcrae2020}. However, accurate measurement of TLS-induced loss in resonators often faces two major challenges. The first challenge is related to the limited drive power one can use to measure the unsaturated resonator loss. Two-level systems are saturable absorbers, and for resonators of small volume such as in planar circuits, in which TLS couple relatively strongly to the internal resonator electromagnetic field, one must measure resonator loss at low intra-resonator photon number, which in many cases is at or below the single-photon limit. Measurement of high-$Q$ microwave resonators at the single-photon level requires the use of an amplifier, the noise of which sets the attainable signal-to-noise ratio (SNR). Ultra-low-noise microwave amplifiers, such as cryogenically-cooled high electron mobility transistor (HEMT) amplifiers or quantum-limited parametric amplifiers~\cite{gao2011}, are typically used in order boost the SNR. More efficient measurement protocols, such as in Ref.~\cite{chen2024}, can also aid in reducing the measurement time. Despite these efforts, measuring the unsaturated TLS absorption in high-$Q$ resonators with high accuracy below the single-photon level remains a challenge.  

The second, more pernicious challenge is how to address temporal fluctuations in the TLS-induced resonator absorption loss \cite{wang2025}. This phenomenon can be explained by the following simplified model. The drive tone, $f_r$, of a resonator interacts with a group of near-resonant TLS whose transition frequencies $f_\mathrm{TLS}$ fall within a window $f_r \pm \delta f$, where $\delta f$ is on the order of the spectral linewidth of the TLS. We can estimate $\delta f$ to be around $1$~MHz, corresponding to the TLS dephasing rate ($1/T_2$) at the milli-Kelvin temperatures typically used when operating microwave superconducting quantum circuits~\cite{Lisenfeld2016}. Crucially, the number of TLS within this $\delta f \sim 1$~MHz interaction zone is not constant in time. Instead, TLS-TLS interactions can cause TLS transition frequencies to drift in and out of this zone, either continuously \cite{Phillips1987} or abruptly (like a telegraph process\cite{Klimov2018}). The fluctuating number of TLS within this $f_r \pm \delta f$ frequency window leads to fluctuations in the TLS absorbed power. Limited by this phenomenon, measurements of TLS-induced absorption loss in quarter-wave resonators in the single photon regime often show uncertainties and fluctuations of over $30\%$ of their mean value \cite{Bejanin2022, Vallieres2024, Earnest2018}.

To reduce the impact of TLS fluctuations on the loss measurement, we propose using multi-wave superconducting resonators. By extending the resonator length from a quarter wavelength ($\lambda/4$) to N/4 wavelengths ($N\lambda/4$), we effectively increase the number of TLS interacting with the microwave tone by a factor of $N$. As a result, the multi-wave resonator design reduces the root-mean-square (RMS) TLS-induced fluctuations in the measurement result by a factor of $\sqrt{N}$ in comparison to a quarter-wave resonator. The longer resonator length also results in lower energy density in the resonator for a given input power. As the absorption rate of any given TLS is proportional to the local electric field intensity, this allows one to measure the TLS-induced loss down to lower TLS saturation levels (1/$N$) without degrading the measurement SNR (i.e., same input power levels).  Finally, the multi-wave resonator exhibits fundamental and harmonic resonances across a wide frequency range that enables broadband studies of frequency-dependent TLS loss.

\section{Device Principle}
\begin{figure}[ht]
    \centering
    \includegraphics[width=\linewidth]{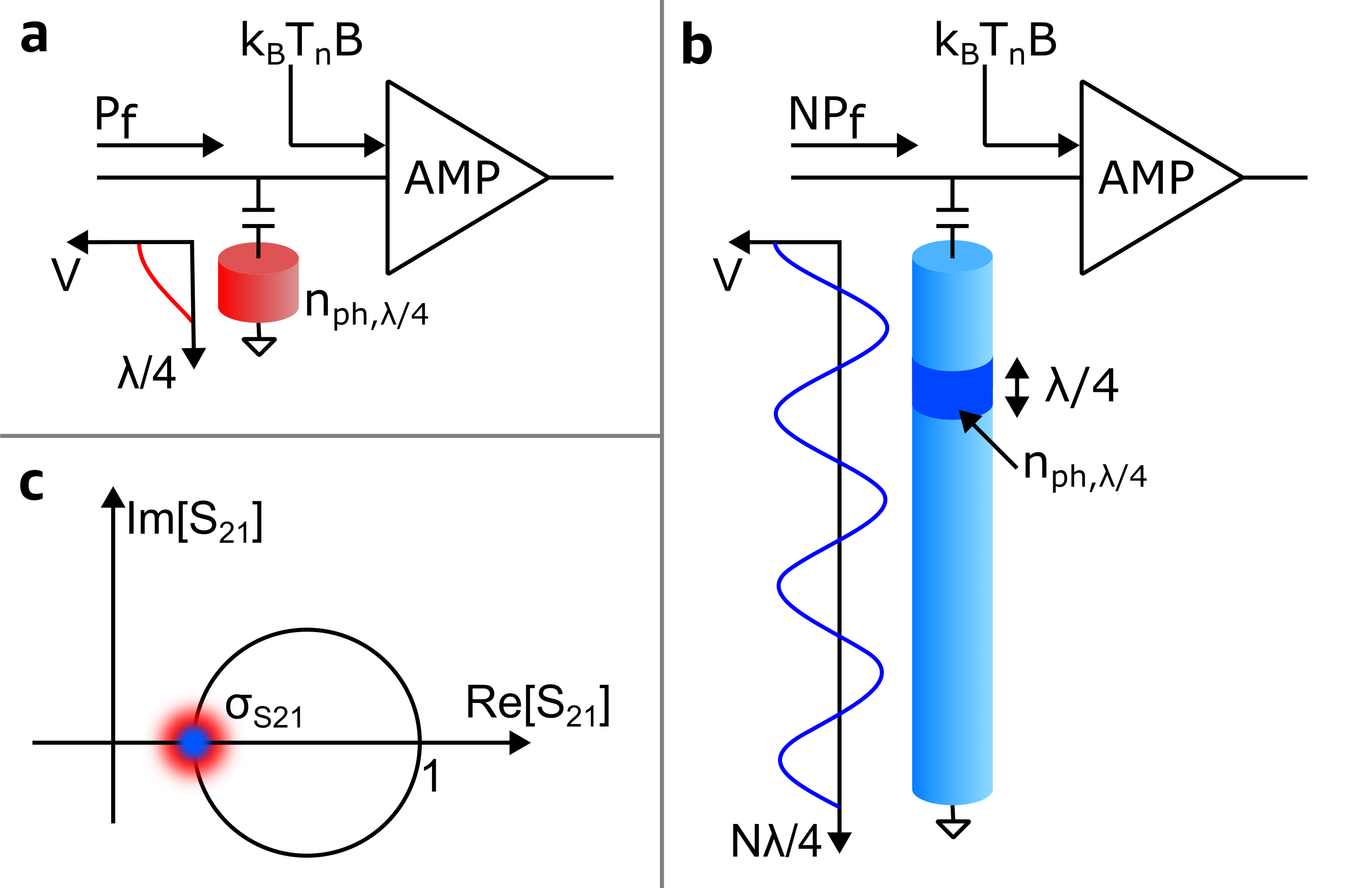}
    \caption{\textbf{Schematic of $\lambda/4$ and $N\lambda/4$ resonators.} \textbf{a.} Schematic of a quarter-wave resonator coupled to a feedline with an amplifier. The resonator is driven with a probe power of $P_f^\mathrm{QW}$ and the output signal is readout through the amplifier whose noise temperature is $T_n$ over the measurement bandwidth $B$. \textbf{b.} Schematic of a multi-wave ($N\lambda/4$) resonator coupled to a feedline operating at the same resonance frequency and with the same amplifier as the \qres resonator in \textbf{a}. The probe power in this case is $P_f^\mathrm{MW} = NP_f^\mathrm{QW}$ so that each \qres section of the multi-wave resonator has the same \nph photons as the quarter-wave. \textbf{c.} Measurement of the $S_{21}$ of the quarter-wave and multi-wave resonators in \textbf{a, b} would yield the same normalized resonance circle. However, the $S_{21}$ measurement uncertainty $\sigma_{S_{21}}$ would be a factor of $\sqrt{N}$ smaller for the multi-wave resonator (blue circle) in comparison to the quarter-wave resonator (red circle). Here we assume the $S_{21}$ uncertainty is dominated by the amplifier noise so that it is given as $\sigma_{S_{21}}\approx \sqrt{k_B T_n B/P_f}$.}    \label{fig:MWPrinciple}
\end{figure}

We explain the principle of the multi-wave resonator design by comparing it to a standard transmission-line quarter-wave resonator, as illustrated in Fig.~\ref{fig:MWPrinciple}. The figure shows circuit diagrams of both a quarter-wave resonator (Fig.~\ref{fig:MWPrinciple}a) and a multi-wave resonator (Fig.~\ref{fig:MWPrinciple}b), each coupled to a feedline that connects to a cryogenic amplifier. Consider exciting the \qres resonator by a microwave tone of power $P_f$ on the feedline which results in an average of \nph photons stored in the resonator. Its \E-field distribution is given by 
\begin{equation}
\vec E(x,y,z) = \vec E_0(x,y)\cos(\frac{2\pi z}{\lambda}),
\label{eqn:DistE}
\end{equation}
where $\vec E_0(x,y)$ represents the \E-field distribution in the x-y cross-section, and $z$ ranges from 0 to $l=\lambda/4$ with $z=0$ representing the open end. For a coplanar waveguide (CPW) resonator, $E_0(x,y)$ corresponds to the the TEM mode of the CPW, which can be determined by solving a 2D electrostatic problem. In the case of a multi-wave resonator, its \E-field distribution maintains the same form as Eqn.~\eqref{eqn:DistE}, but with $z$ extending from $l=0$ to $l=N\lambda/4$, where $N$ is the number of quarter-wavelengths. Importantly, the TLS couple to the local $\vec E$ through their dipole moments which requires $E^\mathrm{MW}(x,y) = E^\mathrm{QW}(x,y)$ in order to have the same TLS-induced loss per unit length, where the superscripts ``MW'' and ``QW'' refer to the multi-wave and quarter-wave cases, respectively. We refer to this requirement as the ``equal \E-field'' condition. Under this condition, each $\lambda/4$ segment of the multi-wave resonator stores an average of \nph photons, matching the quarter-wave resonator, and the total number of photons is $n_\mathrm{ph,N\lambda/4}=N\nph$ because there are $N$ segments. Henceforth, we refer to \nph as the photon ``density'', representing the number of photons per unit $\lambda/4$ section of the resonator.

Now we turn to analyzing the TLS-induced loss in a resonator following the model in Ref.~\cite{Gao2008}. The TLS are assumed to be uniformly distributed within a volume $V_h$, which occupies a portion of the total resonator volume $V$. These TLS contribute to the complex dielectric constant $\epsilon_\mathrm{TLS} = \epsilon_1 - j\epsilon_2$ and loss tangent $\delta_\mathrm{TLS} = \epsilon_2/\epsilon_1$. The mean value of the TLS-induced resonator loss $\xi=1/Q_\mathrm{TLS}$ is given as
\begin{equation}
\xi = \frac{\int_{V_h} \epsilon_2  |\vec{E}|^2 d\vec{r}}{\int_V \epsilon |\vec{E}|^2 \,d\vec{r}}
=F \delta_\mathrm{TLS},~F=\frac{\iint_{\Omega_h}  \epsilon_1  |\vec{E_0}|^2 \,dx\,dy}{\iint_\Omega \epsilon |\vec E_0|^2 \, dx \, dy},
\label{eqn:Qtls_ave}
\end{equation}
where $\epsilon$ is the real part of the dielectric constant while $\Omega_h$ and $\Omega$ represent the TLS occupied area and the total area in the resonator's x-y cross-section, respectively. The TLS filling factor is called $F$ and represents the fraction of the resonator’s total electrical energy stored in the TLS-hosting material. In the derivation of Eqn.~\eqref{eqn:Qtls_ave}, we made use of the \E-field distribution of Eqn.~\eqref{eqn:DistE}. Since $\delta_\mathrm{TLS}$ and $F$ are independent of the resonator length $l$ and the number of quarter-wavelengths $N$, measurements performed on $\qres$ and $\mres$ resonators should yield identical results of $\xi$. Furthermore, according to the standard tunneling TLS model \cite{Phillips1987}, the TLS loss tangent can saturate with temperature and \E-field strength, resulting in a power and temperature-dependent $\xi$ expressed as
\begin{equation}
\xi = \xi_0 \frac{\tanh(\frac{\hbar\omega}{2k_BT})}{\sqrt{1+(|\vec E|/E_c)^2}},~\xi_0 = F\delta_\mathrm{TLS}^0,
\label{eqn:deltaTP}
\end{equation}
where $\omega$ is the angular frequency of the probe tone, $k_B$ is the Boltzmann constant, $T$ is the resonator temperature, $\delta_\mathrm{TLS}^0$ is the intrinsic TLS loss tangent, and $E_c$ is the characteristic electric field strength for TLS saturation. This means that $\xi_0$ is an intrinsic property of the resonator that depends on its geometry and material composition, but is independent of power and temperature. Therefore, it serves as an important figure of merit for comparing different resonator designs and materials. 

Next we consider fluctuations in $\xi$. If the TLS-induced dielectric constant fluctuates on time scales that are much longer than the resonator's period, one would expect to see $\xi$ fluctuations given by
\begin{eqnarray}
\delta \xi(t) &=& \frac{\int_{V_h}  \delta \epsilon_2(\vec r,t) |\vec{E}|^2 \,d\vec{r}}{\int_V \epsilon |\vec{E}|^2 \,d\vec{r}}. \label{eqn:Qtls_t}
\end{eqnarray}
To proceed, we make some assumptions about the correlation function $\langle \delta \epsilon_2(\vec r_1,t_1)\delta \epsilon_2(\vec r_2,t_2) \rangle$ between two regions of $\epsilon_2$. For independently fluctuating TLS, we expect, under the assumption of stationarity, that the correlation function has the form $\langle \delta \epsilon_2(\vec r_1,t_1)\delta \epsilon_2(\vec r_2,t_2) \rangle = C(\vec r_1, t_1-t_2) \delta(\vec r_1 - \vec r_2)$, where $C(\vec r_1, t_1-t_2)$ is some function of a single position coordinate and only the time difference. Furthermore, because the TLS are also assumed to be uniformly distributed in $V_h$, we expect $C(\vec r_1, t_1-t_2)$ to be independent of position $\vec r_1$ which allows it to be rewritten as $C(t_1 - t_2)$. We can now derive the correlation function of the $\xi$-fluctuations to be given as \cite{Gao2008}
\begin{eqnarray}
\langle \delta \xi(t) \delta\xi(0) \rangle&=& \frac{C(t)\int_{V_h}  |\vec{E}|^4 \,d\vec{r}}{\left( \int_V \epsilon |\vec{E}|^2 \,d\vec{r} \right)^2} = C(t)\rho_\mathrm{xy}\rho_z,
\label{eqn:xitxi0}
\end{eqnarray}
where we have used Eqn.~\eqref{eqn:DistE} and the assumption that the TLS distribution is z-independent to write the result as a product of two terms, $\rho_{xy}$ and $\rho_z$. The first term 
\begin{equation}
\rho_\mathrm{xy} = \frac{\iint_{\Omega_h} |\vec{E_0}|^4 \,dx\,dy}{\left(\iint_\Omega \epsilon |\vec E_0|^2 \,dx\,dy\right)^2}
\end{equation}
only involves integrals in the x-y plane and is therefore the same for quarter-wave and multi-wave resonators under the equal \E-field condition. The second term only involves integrals of $z$ which we evaluate to be
\begin{equation}
    \rho_z = \frac{\int_0^{N\lambda/4} \cos^4(\frac{2\pi z}{\lambda})\,dz}{ \left( \int_0^{N\lambda/4} \cos^2(\frac{2\pi z}{\lambda}) \,dz\right)^2} = \frac{6}{N\lambda},~N = 1,3,5,7...
    \label{eqn:rhoz}
\end{equation}
 Due to the $N$ in the denominator, the correlation function is reduced by a factor of $N$ in a multi-wave resonator compared to a quarter-wave resonator. By taking $t=0$ in Eqn.~\eqref{eqn:xitxi0}, we arrive at
 \begin{equation}
 \langle \delta \xi^2 \rangle^\mathrm{MW} = \langle \delta \xi^2 \rangle^\mathrm{QW}/N.
 \label{eqn:factorN}
 \end{equation}
 It follows from the assumption of stationarity and the Wiener-Khinchin theorem that the power spectrum of $\xi$-fluctuations will have the same relationship
 \begin{equation}
 S_{\xi}^\mathrm{MW}(\nu) = S_{\xi}^\mathrm{QW}(\nu)/N.
 \label{eqn:factorNPSD}
 \end{equation}

In addition to reducing the intrinsic TLS-induced fluctuations in $\xi$, a multi-wave resonator has the advantage that it can also improve the measurement SNR for a given level of TLS saturation. Assuming the equal \E-field condition (i.e., equal TLS saturation), equal coupling $Q_c$, and that the internal $Q_i$ is limited by $Q_\mathrm{TLS}$, the two types of resonators should have the same $Q_i$ and loaded $Q$. This indicates that an $S_{21}$ measurement of both resonators would yield the same resonance circle. However, the probe power used to measure the $\mres$ resonator would be N times greater, $P_{f}^\mathrm{MW} = N P_{f}^\mathrm{QW}$, despite the same $S_{21}$ response. This follows from the fact that the equal \E-field condition is equivalent to having the same internal microwave power $P_\mathrm{int}$, where $P_\mathrm{int} = \frac{2}{N\pi}\frac{Q^2}{Q_c}P_f$ \cite{Gao2008}. Therefore, assuming the same amplifier noise level, the $S_{21}$ measurement uncertainty $\sigma_{S_{21}}$ is expected to be a factor of $\sqrt{N}$ smaller for the multi-wave resonator in comparison to the quarter-wave resonator as we show in Fig.~\ref{fig:MWPrinciple}c. Conversely, this implies that when $P_{f}^\mathrm{MW} = P_{f}^\mathrm{QW}$, then $E^\mathrm{MW}(x,y) \sim E^\mathrm{QW}(x,y)/\sqrt{N}$. This indicates that the multi-wave resonator can reduce the intra-resonator energy density $|E(x,y)|^2$ by a factor of N while maintaining a similar SNR as that of a quarter-wave resonator, thus making observation of $\xi_0$ easier. As a summary of this section, we list the relationships between the multi-wave and quarter-wave versions of various parameters below along with their definitions in Table~\ref{tab:rmwqw}.

\begin{align*}
  \left.\begin{array}{r@{\mskip\thickmuskip}l}
    \vec{E}^\mathrm{MW} &= \vec{E}^\mathrm{QW} \\[1em]
    f_{r, N}^\mathrm{MW} &= f_{r, 1}^\mathrm{QW} \\[1em]
    Q_c^\mathrm{MW} &= Q_c^\mathrm{QW} \\[1em]
  \end{array} \right\}
  \quad \implies \quad
  \left\{\begin{array}{r@{\mskip\thickmuskip}l}
  n^\mathrm{MW}_{\text{ph}} &= N n^\mathrm{QW}_{\text{ph}} \\[1em]
  \nph^\mathrm{MW} &= \nph^\mathrm{QW}\\[1em]
    P_f^\mathrm{MW} &= NP_f^\mathrm{QW} \\[1em]
    \langle \delta \xi^2 \rangle^\mathrm{MW} &= \langle \delta \xi^2 \rangle^\mathrm{QW}/N \\[1em]
    S_\xi^\mathrm{MW} &= S_\xi^\mathrm{QW}/N \\[1em]
    [\sigma_{S_{21}}^2]^\mathrm{MW} &= [\sigma_{S_{21}}^2]^\mathrm{QW}/N
  \end{array}\right.
\end{align*}

\begin{table}[!h]
    \centering
    \caption{Resonator parameter definitions}
    \begin{ruledtabular}
    \begin{tabular}{cl}
         Parameter & Definition\\
        \specialrule{.5pt}{2pt}{2pt}
        $\vec{E}$ & Local electric field\\
        $f_{r, N}$ & Nth harmonic resonance frequency \\
        $Q_c$ & Coupling quality factor \\
        $n_{ph}$ & Average total photon number \\ 
        \nph & Average photon number per $\lambda/4$ length \\
        $P_f$ & Feedline power \\
        $\langle \delta \xi^2 \rangle$ & Variance in TLS-induced loss \\
        $\sigma_{S_{21}}^2$ & Variance in $S_{21}$ \\
        $S_\xi$ & Power spectral density of TLS-induced loss
    \end{tabular}
    \end{ruledtabular}
    \label{tab:rmwqw}
\end{table}

\section{Device fabrication and experimental setup}
\begin{figure}[tb]
    \centering
    \includegraphics[width=\linewidth]{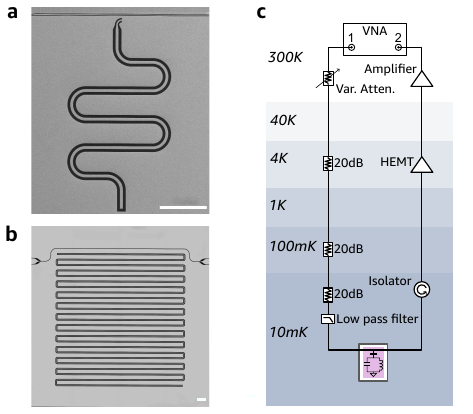}
    \caption{\textbf{Devices and experimental setup.} White scale bars at the bottom right of each image correspond to \SI{500}{\micro\meter}. \textbf{a.} Scanning electron micrograph of the $\qres$ resonator design. \textbf{b.} False-color optical micrograph of the $\mres$ resonator design. \textbf{c}. Measurement setup used in this study.} 
    \label{fig:1}
\end{figure}
To compare the performance of the $\mres$ design against the standard $\qres$ design, we fabricated two devices from a \SI{100}{\nano\meter} thick Al film deposited on a high-resistivity ($~10$~kOhm-cm) silicon substrate. One device contained eleven hanger-style $\qres$ resonators coupled to a common transmission line while the other contained a single $\mres$ resonator with a fundamental frequency of \SI{165}{\MHz} as shown in Fig.~\ref{fig:1}a,b. Both resonator designs used a coplanar-waveguide (CPW) geometry with \SI{30}{\micro\meter} center strip and gap widths. We chose adjacent dies for each device in order to help eliminate spatial variations during fabrication.

The devices were cooled in a dilution refrigerator and $S_{21}$ measurements on each device were carried out using a vector network analyzer (VNA) in a setup shown in Fig.~\ref{fig:1}c. To best utilize the measurement time, we implemented the hybrid approach of parameter-constrained conformal mapping method as described in Ref \cite{chen2024}. At high powers ($\nph \gtrsim 10^4$) in the linear regime, we performed frequency sweeps of the $S_{21}$ measurement centered at $f = f_r$ and extracted the loss along with power-independent resonator parameters by fitting the data using the diameter correction method \cite{Khalil2012}; the error bars quote the $1\sigma$ confidence interval. At lower powers, we measured the time-dependent $S_{21}(f, t)$ on resonance at a single frequency $f=f_r$ for a few seconds. We then conformally mapped the $S_{21}(f_r, t)$ time-series into $1/Q_i(t)$ using the power-independent parameters determined at high power to obtain the average $1/Q_i$ value at each lower power \cite{chen2024}; in this case, the error bars quote the standard error of the mean. Finally, the TLS-induced loss was obtained by subtracting the power-independent loss component ($1/Q_\mathrm{i,hp}$, measured at highest power) from the total resonator loss ($1/Q_i$), yielding $\xi = 1/Q_\mathrm{TLS} = 1/Q_i - 1/Q_\mathrm{i,hp}$ (see supplementary material for more details on device design, fabrication and loss measurement method).

\section{Results}

\subsection{Comparison of $\xi$ versus \nph for multi-wave and quarter-wave resonators}
\begin{figure}[tb]
    \centering
    \includegraphics[width=\linewidth]{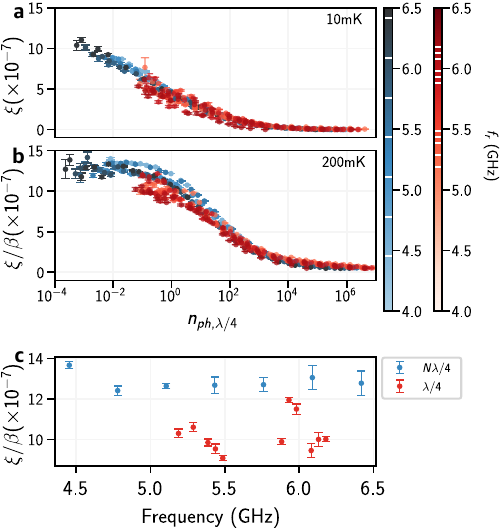}
    \caption{\textbf{Measurement of $\xi$ versus \nph for Al-based $\qres$ and $\mres$ CPW resonators.} \textbf{a.} $\xi$ vs \nph measured at \SI{10}{\milli\kelvin} for eleven $\qres$ resonators (red) and seven $\mres$ resonances (blue) covering a resonance frequency range of \qtyrange[range-phrase = ~--~]{4}{6.5}{\GHz}. Corresponding resonance frequencies are indicated by the white marks in the color bar. \textbf{b.} Temperature-corrected $\xi$ vs \nph for the same resonators in \textbf{a} measured at \SI{200}{\milli\kelvin}. The temperature factor of $\beta = \tanh(hf_r/2k_B T)$, as defined in Eqn.~\eqref{eqn:deltaTP}, where $T = \SI{0.2}{\kelvin}$ has been divided out. The quasi-particle loss from the elevated temperature has also been removed. \textbf{c.} Temperature-corrected loss values extracted from \textbf{b} at the lowest photon densities plotted against $f_r$.} 
    \label{fig:3}
\end{figure}

\begin{figure*}[tb]
    \centering
    \includegraphics[width=\linewidth]{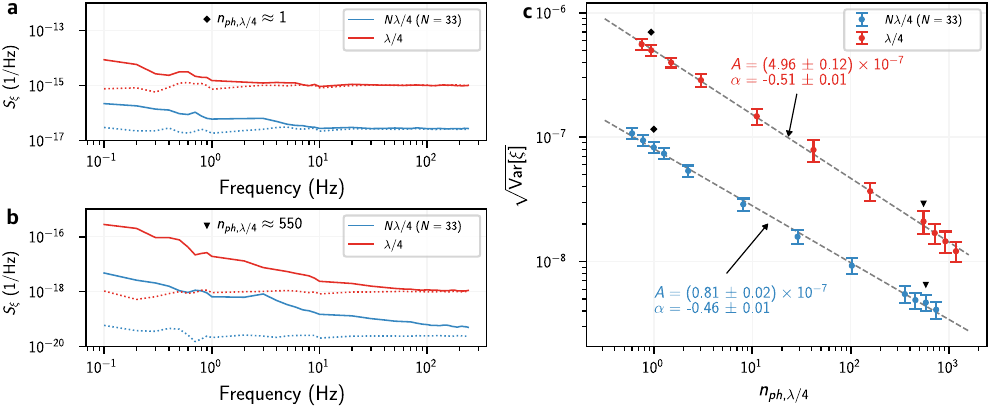}
    \caption{\textbf{Power spectral densities and RMS fluctuations of $\xi$ against \nph for $\qres$ and $\mres\;(N=33)$ resonators measured at \SI{10}{\milli\kelvin}.} \textbf{a.} Solid lines correspond to the power spectral densities of the data projected along the $\xi$ quadrature at $\nph \approx 1$. Dotted lines correspond to the power spectral densities of the transmission background taken off resonance ($f = f_r + \SI{2}{\MHz}$) after projection. \textbf{b.} The same as in \textbf{a} but for $\nph \approx 550$. \textbf{c.} Each point corresponds to data taken on-resonance for \SI{100}{\second} at a sample rate of \SI{500}{\Hz}. The data was subsequently projected along the dissipation quadrature $\xi$ from which the PSD was estimated using Welch's method. Next, the variance was calculated by integrating the PSD down to \SI{0.1}{\Hz}. The gray dashed lines indicate best fits to a scaling law of the form $\sqrt{\mathrm{Var}[\xi]} = A\nph^\alpha$. The points marked with a $\blacklozenge$ or $\blacktriangledown$ have their corresponding PSDs shown in \textbf{a} and \textbf{b}, respectively.} 
    \label{fig:5}
\end{figure*}

We experimentally compare the power-dependent TLS loss $\xi$ measured from the $\mres$ and $\qres$ resonators in Fig.~\ref{fig:3}. Measurements were first performed at the base temperature \SI{10}{\milli\kelvin} of the dilution fridge. The loss curves from the $\qres$ resonators (red) and the $\mres$ resonator (blue), with $29 \leq N \leq 37$, lie roughly on top of each other when plotted against photon density \nph. This overlap confirms that both resonator types yield the same measurement result of $\xi$, as expected given their identical cross-sectional CPW geometry and fabrication. The average loss across all frequencies at $\nph \approx 1$ is $\xi = (4.66 \pm 0.49) \times 10^{-7}$ ($ (3.96 \pm .57) \times 10^{-7}$) for $\mres$ resonances ($\qres$ resonators). Notably, the loss curves do not plateau for either type of resonator even at the lowest power, making it impossible to extract the intrinsic loss $\xi_0=F\delta_0$.

To address this issue, we repeated the measurements at a temperature of \SI{200}{\milli\kelvin}. Raising the temperature extends the plateau region to higher powers by increasing the critical field $\vec{E}_c$ that results in TLS saturation. This occurs because the critical field scales as $|\vec{E}_c| \propto (T_1T_2)^{-1/2}$, where both relaxation ($1/T_1$) and dephasing ($1/T_2$) rates increase with temperature \cite{Gao2008, Phillips1981, Faoro2015}. In order to remove the temperature-dependent factor ($\beta = \tanh(hf_r/2k_B T)$) out of the TLS loss in Eqn.~\eqref{eqn:deltaTP}, we plot in Fig.~\ref{fig:3}b a temperature-corrected loss of $\xi/\beta$.  The thermal quasi-particle loss associated with the elevated temperature \cite{Gao2008, Gao2008_2} has been accounted for and subtracted off by the power independent loss term $1/Q_\mathrm{i,hp}$. A distinct plateau region is observed in the $\mres$ curves at low photon densities ($\nph < 0.1$), while the $\qres$ curves follow a similar curve shape but are limited to an $N$ times higher region of \nph due to the limited SNR at the lowest \nph. Reading directly from this plateau region of the \mres resonator, where the TLS are in their ground state, we can safely determine the intrinsic loss parameter $\xi_0 = F\delta_\mathrm{TLS}^0 = 1.25\times 10^{-6}$. 
 
In Fig.~\ref{fig:3}c, we plot $\xi$ against $f_r$ for each of the resonators. The loss values were extracted from the temperature-corrected \SI{200}{\milli\kelvin} data at the lowest measured photon density \nph. The $\qres$ resonator loss values (red) are slightly lower (by $15\%$) than the $\mres$ resonator, primarily because the $\qres$ resonator measurements barely reach the plateau region. We observe no clear frequency dependence of the loss across a span of $\sim$\SI{2}{\GHz}, consistent with the standard model of a uniform TLS density of states in this frequency range \cite{Phillips1987, Martinis2013}.

\subsection{Comparison of $\xi$-fluctuations measured for multi-wave and quarter-wave resonators}

\begin{figure}[tb]
    \centering
    \includegraphics[width=\linewidth]{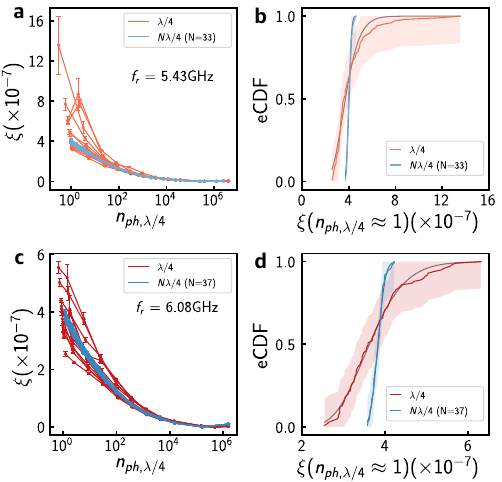}
    \caption{\textbf{Repeated $\xi$ measurements $\qres$ and $\mres$ resonators at \SI{10}{\milli\kelvin} over \SI{12}{\hour}}. \textbf{a,c.} show the repeated measurements of $\xi$ against \nph over \SI{12}{\hour} for two $\qres$ resonators (red) and two $\mres$ resonances (blue) at \SI{5.43}{\GHz} and \SI{6.08}{\GHz}. For clarity, we only plot 12 randomly selected curves out of the total 100 curves for each resonator. \textbf{b,d.} Empirical cumulative distribution functions of $\xi(\nph \approx 1)$ taken from the 100 curve data sets. Gray lines indicate best fits to a log-normal distribution. The filled areas correspond to the Dvoretzky–Kiefer–Wolfowitz confidence bounds calculated at the 99\% confidence level.} 
    \label{fig:6}
\end{figure}

Based on the presented analysis, a multi-wave resonator is expected to show an $\sqrt{N}$-fold reduction in $\xi$-fluctuations compared to a quarter-wave resonator when operated at similar resonance frequencies and excited to equivalent photon densities \nph. We experimentally demonstrate this insensitivity to TLS drift on both short (seconds) and long (hours) time scales. 

First, we show the $\sqrt{N}$-fold improvement on a short time scale of $0.2-10$~s. Note that on these timescales the noise in the inferred value of $\xi$ is a combination of both the white noise of the amplifier and the low-frequency noise of the TLS frequency fluctuations. We compare the power spectral densities (PSDs) between a $\mres$ resonator with $N=33$ (\SI{5.43}{\GHz}) and a $\qres$ (\SI{5.23}{\GHz}) resonator. The raw data was obtained by measuring the on-resonance time-dependent $S_{21}(f, t)$ at a single frequency $f\approx f_r$ for \SI{100}{\second} at a sample rate of \SI{500}{\Hz}. We subsequently projected the data onto the dissipation quadrature $\hat \xi$ and converted it to $\xi(t)$, from which we computed the PSD $S_\xi(\nu)$ using Welch's method \cite{welch}. The \SI{100}{\second} length was chosen to reduce the uncertainty in the PSD estimate at low frequencies near \SI{0.1}{\Hz}. This analysis protocol follows standard procedures for characterizing quadrature-dependent resonator noise \cite{Gao2008} and more details are included in the the supplementary material. 

We plot the resulting $S_\xi(\nu)$ as solid lines in Fig.~\ref{fig:5}a and Fig.~\ref{fig:5}b  for $\nph \approx 1$ and $\nph \approx 550$, respectively. For reference, we also include the system noise floors (dashed lines) measured off-resonance at $f = f_r + \SI{2}{\MHz}$. For both resonators, the noise floors appear as white noise which converge with the on-resonance PSDs (solid lines) for $\nu > \SI{10}{\Hz}$. Towards lower frequencies ($\nu < \SI{1}{\Hz}$), $S_\xi(\nu)$ exhibits a 1/f-noise behavior that rises above the white noise floor (set by the amplifier noise) which suggests that TLS-induced fluctuations are the dominant noise mechanism in this frequency range. As discussed previously, the noise power of both the TLS frequency fluctuations and the amplifer noise are expected to reduce by a factor of $N$ in the multi-wave resonators. Indeed, we observe a large separation between the two curves noise power spectral densities of $S_\xi(\nu)$ throughout the entire frequency range, with the $\mres$ resonator being more than 30 times lower than the $\qres$ resonator. This result demonstrates the effectiveness of the multi-wave resonator design in substantially reducing the $\xi$-fluctuation for short time scales $<$\SI{10}{\second}. The amount of noise power reduction is also in good agreement with the prediction of a factor N = 33 (Eqn.~\ref{eqn:factorNPSD}). 

We further demonstrate this reduction over a range of microwave signal powers by plotting the RMS fluctuations in $\xi$, $\sqrt{\mathrm{Var}[\xi]}$, against \nph in Fig.~\ref{fig:5}c. Each point is calculated by integrating its corresponding PSD down to \SI{0.1}{\Hz}. The gray dotted lines correspond to best fits to a scaling law of the form $\sqrt{\mathrm{Var}[\xi]} = A\nph^\alpha$. Comparing the pre-factors indicates that $A_{\qres}/A_{\mres} = 6.2 \pm 0.2$ showing again a reduction in noise on the order of $\sqrt{N}$ across a wide range of photon densities. We note that the reduction factor of exactly $\sqrt{N}$ is derived under the assumption that both resonators share identical geometry, $Q_c$, $Q$, and are excited to produce equivalent E-field strength and distribution $\vec E_0(x,y)$. In our case, $Q_c$ for the two resonators were chosen to be close but still differed by $15\%$. Additionally, imperfection in power calibration and chip-to-chip variations in TLS density may also affect the actual observed $S_\xi(\nu)$ reduction factor. 
 
Next we compare the long-term performance on a time scale of hours. We conducted repeat measurements of $\xi$ versus \nph curves (similar to those in Fig.~\ref{fig:3}) at \SI{10}{\milli\kelvin} over a \SI{12}{\hour} period on two $\mres$ resonances and two $\qres$ resonators with similar resonance frequencies at \SI{5.43}{\GHz} and \SI{6.08}{\GHz}. Each $\xi$ versus \nph curve took about \SI{2}{\minute} (12 power steps with $10$~s measurement time at each step) to collect and a total 100 curves were collected per resonator/resonance. We note that the multi-wave and quarter-wave data sets were taken during separate \SI{12}{\hour} measurement periods. 

We pick twelve random $\xi$ versus \nph curves from this \SI{12}{\hour} period for each resonator type and overlay them in Fig.~\ref{fig:6}a and c. Qualitatively, the multi-wave curves show notably less spread than the $\qres$ curves over the \SI{12}{\hour}. For numerical comparison, we calculated the empirical cumulative distribution functions (eCDF) of the $\xi$ values measured for $\nph \approx 1$ from the full 100 curve data sets and plot them in Fig.~\ref{fig:6}b and d; the corresponding Dvoretzky–Kiefer–Wolfowitz confidence bands at the 99\% level are shown as the filled areas \cite{DKW}. These eCDFs were fitted to a log-normal distribution and we found good agreement; the results of the fits are tabulated in Table~\ref{table}. Compared to their $\qres$ counterparts, the $\mres$ resonances demonstrate reductions in the RMS fluctuation of $\xi$ by factors of 5.5 and 4.9 for $N=33$ and $N=37$, respectively. This corresponds to a five-fold decrease in the relative uncertainties of $\xi$ from $\gtrsim20\%$ to below $\lesssim5\%$, which is achieved by only \SI{10}{\second} of measurement time. The smaller improvement of the $N=37$ resonance over its corresponding \qres resonator could attributed to potential intra-chip fabrication variations; we have observed before that different resonators within the same die can have different levels of loss and fluctuations. These findings further confirm the improved performance of the $\mres$ design over the standard $\qres$ design, demonstrating reduced $\xi$-fluctuations and better measurement sensitivity across both short and long time scales.

\begin{table}[!h]
    \centering
    \caption{Mean $\mathrm{E}[\xi]$, variance $\mathrm{Var}[\xi]$, and relative errors $\sqrt{\mathrm{Var}[\xi]}/\mathrm{E}[\xi]$ obtained from fitting the empirical cumulative distributions of $\xi_{1ph}$ in Fig.~\ref{fig:6} to a log-normal distribution.}
    \begin{ruledtabular}
    \begin{tabular}{ccccc}
         $f_r$ (GHz) & N & $\mathrm{E}[\xi]/10^{-7}$ & $\mathrm{Var}[\xi]/10^{-14}$ & $\sqrt{\mathrm{Var}[\xi]}/\mathrm{E}[\xi]$\\
        \specialrule{.5pt}{2pt}{2pt}
        5.43 & 1 & $3.83 \pm 0.016$ & $0.968 \pm 0.055$ & 0.256\\
        5.43 & 33 & $4.018 \pm 0.001$ & $0.032 \pm 0.0009$ & 0.044\\
        6.08 & 1 & $3.677 \pm 0.005$ & $0.476 \pm 0.012$ & 0.188\\
        6.09 & 37 & $3.826 \pm 0.001$ & $0.019 \pm 0.0005$ & 0.036\\
    \end{tabular}
    \end{ruledtabular}
    \label{table}
\end{table}

\section{Discussion}
The multi-wave resonator device significantly improves TLS loss measurement sensitivity, enabling us to investigate deep into the sub-single-photon regime. The results shown in Fig.~\ref{fig:3} highlight an important distinction between two key quantities, $\xi_{1ph}$ and $\xi_0$. The quantity $\xi_{1ph}$ characterizes the TLS loss near $\nph \approx 1$ and is dependent on circuit operating conditions (such as input power and its proximity to the TLS-saturation power) while $\xi_0$ represents the intrinsic TLS loss limited by material and circuit design parameters. Importantly, these two values, $\xi_{1ph}$ and $\xi_0$, can differ due to TLS saturation effects. 

In our \SI{10}{\milli\kelvin} measurements, neither resonator type exhibits a loss plateau at the lowest powers, despite reaching photon densities as low as $\nph \approx 10^{-1}$ in the $\qres$ resonator case and $\nph \approx 10^{-3}$ in the $\mres$ resonator case. Given the absence of a plateau, $\xi_0$ cannot be extracted from the \SI{10}{\milli\kelvin} measurements. Conversely, at an elevated temperature of \SI{200}{\milli\kelvin}, a distinct plateau emerges in the loss curves, allowing us to accurately determine $\xi_0$. Notably, $\xi_{1ph}$ is approximately half the value of $\xi_0$ in the resonators studied here. This demonstrates that combining temperature control with the multi-wave resonator can provide a more comprehensive characterization of TLS-induced loss and lead to better comparisons between different materials and designs~\cite{Earnest2018, Vallieres2024}.

It should also be noted that measuring a single \mres resonator is, in principle, equivalent to measuring $N$ identical \qres resonators at the same resonance frequency and averaging the result. However, the multi-$\lambda/4$-resonator approach is significantly more demanding from a resource standpoint, either requiring more hardware if parallelized or more time if performed sequentially. In addition, the improvements of the multi-wave resonator approach extend to measurements of the resonator frequency as well, thereby enabling accurate measurements of the temperature-dependent frequency shift due to TLS.

Finally, we mention some considerations that should be taken into account when utilizing an $\mres$ resonator to probe absorption loss. First, since the coupling quality factor $Q_c$ varies with frequency, the coupling structure must be carefully designed to accommodate the specific frequency band of interest. Second, the larger physical footprint of the resonator means that is is more susceptible to parasitic modes in comparison to the standard \qres resonator. Airbridges or wire-bonds should be added to the device to maintain ground plane continuity and suppress unwanted modes. 

\section{Conclusion}
In conclusion, we have shown that a multi-wave resonator device reduces the TLS-induced loss measurement uncertainty in comparison to conventional quarter-wave resonator devices. Our theoretical and experimental results demonstrate that the multi-wave resonator reduces fluctuations from both the TLS in the device and the cryogenic amplifier, resulting in a $\sqrt{N}$ reduction in measurement uncertainty, both on short time scales (seconds) and long time scales (hours). Additionally, the multi-wave resonator reduces the photon density in the resonator for a given SNR, enabling more efficient observation of the fully-unsaturated TLS contribution to the resonator loss, $\xi_0$. Finally, the multiple resonances of a multi-wave resonator provides a spectral analysis of the TLS-induced absorption loss. Taken together, these properties of a multi-wave resonator make it a valuable tool for investigating the impacts of materials, resonator designs, and fabrication processes on quantum device performance.

\section{Acknowledgments}

We thank the staff from across the AWS Center for Quantum Computing (CQC) that enabled this project. We also thank \mbox{Simone Severini}, \mbox{James Hamilton}, \mbox{Nafea Bshara}, and \mbox{Peter DeSantis} for their support of the research activities at the CQC.

\section{Data availability}
Data is available from the authors upon reasonable request.

\bibliographystyle{apsrev.bst}
\bibliography{refs}

\clearpage
\appendix

\onecolumngrid
\begin{center}
{\large \textbf{Supplementary Information for ``Broadband and high-precision two-level system loss measurement using superconducting multi-wave resonators"} \par}
\end{center}
\twocolumngrid

\section{Device design and fabrication}
\begin{figure}[h]
    \centering
    \includegraphics[width=\linewidth]{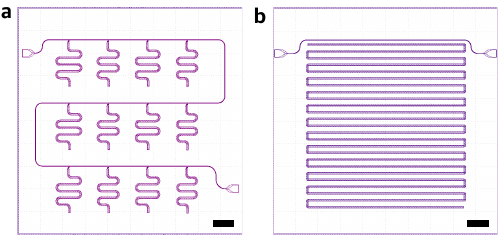}
    \caption{\textbf{a} Quarter-wave device design. \textbf{b} Multi-wave device design. The black scale bar at the bottom right of each image corresponds to \SI{1}{\milli\meter}.} 
    \label{fig:MWQWDesign}
\end{figure}

One \SI{1}{\centi\meter}$\times$\SI{1}{\centi\meter} quarter-wave device (Fig.~\ref{fig:MWQWDesign}a) and one \SI{1}{\centi\meter}$\times$\SI{1}{\centi\meter} multi-wave device (Fig.~\ref{fig:MWQWDesign}b) were fabricated on the same high-resistivity silicon wafer deposited with \SI{100}{\nano\meter} thick aluminum. All resonators used a coplanar-waveguide geometry with \SI{30}{\micro\meter} center strip and \SI{30}{\micro\meter} gap widths. The quarter-wave device contained twelve \qres resonators inductively coupled to a common feedline in a hanger-style configuration. The multi-wave device contained a single resonator of length \SI{18.1}{\centi\meter} capacitively coupled to a feedline. We note that the choice of inductive versus capacitive coupling does not affect the results obtained in this study.
 
\section{Nominal resonator parameters}

\begin{figure}[h]
    \centering
    \includegraphics[width=\linewidth]{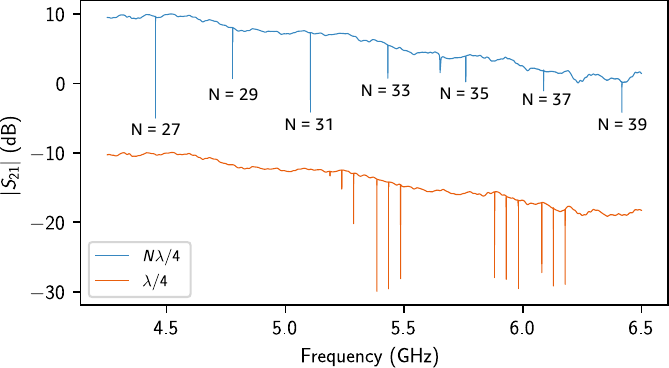}
    \caption{Broadband $S_{21}$ versus frequency of the \qres (orange) and \mres (blue) resonator devices.} 
    \label{fig:wideband}
\end{figure}

Resonator quality factors were determined from measurements of the $S_{21}$ scattering parameter of each device. These measurements were carried out using a VNA in a dilution fridge setup as described in Fig.~2c of the main text. Resonances for the \qres device (orange) and the \mres device (blue) were identified using a broadband frequency sweep between \SI{4.25}{\GHz} and \SI{6.5}{\GHz} as shown in Fig.~\ref{fig:wideband}; the traces have been vertically shifted for clarity. 

For the \qres device, we identified twelve resonances dips in accordance with the 12 resonators on the device. We found that one resonator at \SI{5.19}{\GHz} showed unusually high power-independent loss which attributed to the presence of trapped flux in the ground plane near the resonator. This resonator has been excluded from the study. 

For the \mres device, we identified seven resonances with a frequency spacing of \SI{0.33}{\GHz} within the \qtyrange[range-phrase = ~--~]{4.25}{6.5}{\GHz} range in accordance with the design. We also picked up one spurious resonance at \SI{5.65}{\GHz} whose resonance frequency does not match the expected harmonics of the \mres design. The presence of this extra resonance is not unexpected given that the \mres resonator is more susceptible to parasitic modes due to the large division of the ground plane. A table of the resonator quality factors extracted at high photon densities ($\nph > 10^4$) for both devices can be found below. 

\begin{table}[!h]
    \centering
    \caption{Quarter-wave and multi-wave resonator frequencies and quality factors at $\nph > 10^4$}
    \begin{ruledtabular}
    \begin{tabular}{ccccc}
        Type & N & $f_r$ (GHz) & $Q_i$ ($\times 10^6)$ & $Q_c$ ($\times 10^6$)\\
        \specialrule{.5pt}{2pt}{2pt}
        MW & 27 & 4.45 & 9.46 & 2.42 \\
        MW & 29 & 4.77 & 2.77 & 2.19 \\
        MW & 31 & 510 & 5.42 & 2.27 \\
        MW & 33 & 5.43 & 3.33 & 2.34 \\
        MW & 35 & 5.76 & 1.63 & 3.31 \\
        MW & 37 & 6.08 & 2.20 & 5.40 \\
        MW & 39 & 6.42 & 3.89 & 6.58 \\\cmidrule[0.2pt](l{0.5em}r{1.5em}){1-5}
        QW & 1 & 5.23 & 1.94 & 3.31 \\
        QW & 1 & 5.28 & 2.71 & 2.25 \\
        QW & 1 & 5.39 & 7.53 & 1.79 \\
        QW & 1 & 5.43 & 8.37 & 2.06 \\
        QW & 1 & 5.48 & 6.50 & 1.81 \\
        QW & 1 & 5.88 & 7.05 & 2.12 \\
        QW & 1 & 5.93 & 6.83 & 2.04 \\
        QW & 1 & 5.98 & 6.61 & 1.83 \\
        QW & 1 & 6.08 & 4.32 & 2.10 \\
        QW & 1 & 6.13 & 5.67 & 2.48 \\
        QW & 1 & 6.18 & 3.97 & 2.06
    \end{tabular}
    \end{ruledtabular}
\end{table}

\section{Loss versus power measurement and data analysis}

\begin{figure*}[tbh]
    \centering
    \includegraphics[width=1\linewidth]{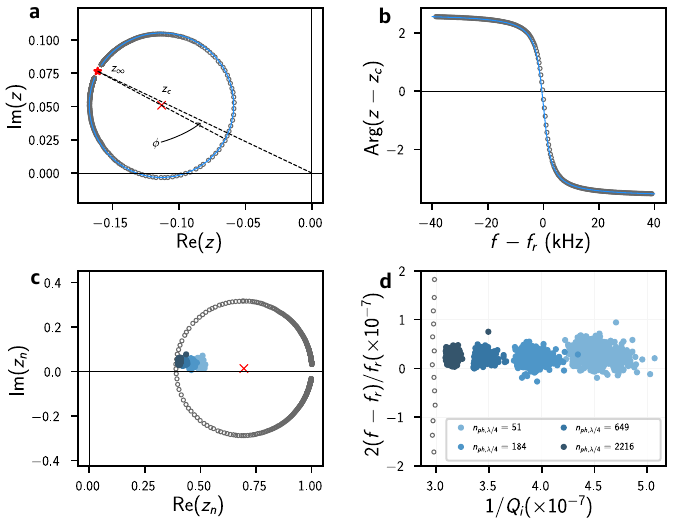}
    \caption{\textbf{Method for extracting the power-dependent loss.} The data presented here was taken on the $N = 33$ harmonic of the \mres resonator device. \textbf{a.} The unwrapped, frequency-dependent data $z$ (grey circles) taken at high photon density ($\nph \approx 7000$) is plotted alongside a best fit (blue line) using the diameter correction method. Other resonator parameters used in describing the resonance circle are also shown. \textbf{b.} Frequency dependent phase of the resonance circle (gray circles) plotted alongside a best fit (blue line) to a function of the form $\mathrm{Arg}(z-z_c) = \theta + 2\arctan(2Q_l(f-f_r)/f_r)$. \textbf{c.} The normalized, high photon density resonance circle (gray circles) $z_n$ plotted alongside the low photon density, single frequency data taken on-resonance (blue dots). \textbf{d.} Using the power-independent resonator parameters determined at high photon density, the data shown in \textbf{c.} is conformally mapped to directly obtain the resonator loss $1/Q_i$ and fractional frequency shift $2(f-f_r)/f_r$. 
    }
    \label{fig:lossanalysis}
\end{figure*}

We used a hybrid approach to extract the power-dependent resonator loss $\xi$. All data presented in this section was taken on the $N=33$ harmonic of the \mres resonator device. We assume the $S_{21}$ data can be modeled as
\begin{equation}
    S_{21}(f) = e^{-j2\pi f \tau}z = e^{-j2\pi f \tau}z_\infty \frac{Q_l/Q_c e^{j\phi}}{1+2Q_l(f-f_r)/f_r},
    \label{eqn:s21}
\end{equation}

where $\tau$ is the cable delay including the delay on the chip, $z_\infty$ is the off-resonance point, $\phi$ is a mismatch angle associated with impedance mismatches in the signal chain, $f_r$ is the resonance frequency, and $Q_l$, $Q_c$ are the loaded and coupling quality factors respectively. The intrinsic quality factor $Q_i$ and loss $\xi$ are subsequently calculated as 

\begin{align}
    \frac{1}{Q_i} &= \frac{1}{Q_l} - \frac{\cos\phi}{Q_c} \\[1em]
    \xi &= \frac{1}{Q_i} - \frac{1}{Q_i^*},
\end{align}

where $Q_i^*$ is the $Q_i$ value obtained in the power-independent regime at high photon densities and represents the TLS-independent loss channels. Extracting $\xi$ at high photon densities can be done by fitting the data to Eqn.~\ref{eqn:s21} using the conventional diameter correction method \cite{Khalil2012}. We showcase an example DCM fit (blue line) of data (gray circles) taken at $\nph \approx 7000$ in Fig.~\ref{fig:lossanalysis}a; note that we are plotting the unwrapped data $z = e^{j2\pi f \tau}S_{21}(f)$. As part of the fitting procedure, a non-linear fit of the phase $\mathrm{Arg}(z-z_c)$ to the function $2\arctan(2Q_l(f-f_r)/f_r) + \theta$ is performed, where $z_c$ is the resonance circle center, $f_r$ is the resonance frequency, and $Q_l$ is the loaded quality factor. We show the results of this phase fit in Fig.~\ref{fig:lossanalysis}b. 

At low photon densities approaching $\nph \lesssim 1$, the SNR is reduced and the noisy data can cause the DCM fitting procedure to produce unreliable estimates of $\xi$. To mitigate this issue, we utilized the strategy outlined in Ref \cite{chen2024}. We collected $S_{21}$ data at varying \nph at a single frequency $f \approx f_r$. Next, we normalized the data by the off-resonance point to obtain $z_n = z/z_\infty$ which we plot as blue dots in Fig.~\ref{fig:lossanalysis}c; the normalized version of the resonance circle from Fig.~\ref{fig:lossanalysis}a (gray circles) is also shown for reference. We proceeded to conformally map the normalized data $z_n$ to obtain a new set of values $w$ whose real component $\mathrm{Re}[w]$ corresponded to estimates of $1/Q_i$ as we show in Fig.~\ref{fig:lossanalysis}d. The final $1/Q_i$ estimate and error were reported by taking the mean and standard error of $\mathrm{Re}[w]$ which was then converted into a final estimate of $\xi$. This same procedure was used for the measurements taken at \SI{200}{\milli\kelvin}.

\section{Noise power spectral density measurement and data analysis}

We obtained the raw data for the power spectral densities (PSD) shown in Fig.~4a,b of the main text by measuring the time-dependent $S_{21}(f, t)$ at a single frequency $f\approx f_r$ for \SI{100}{\second} at a sample rate of \SI{500}{\Hz}. To calculate the PSD in $\xi$, we processed the data as follows. First, we factored out the cable delay from the $S_{21}$ to obtain the unwrapped data $z$ and then subtracted the time-averaged mean $\langle z \rangle_t$ to obtain $\delta z = z - \langle z \rangle_t$. We then projected the $\delta z$ values along two constant directions,

\begin{align}
    \hat r &\propto \frac{\partial z}{\partial (1/Q_i)}\Bigr\rvert_{f = f_r} \\[1em]
    \hat \theta &\propto \frac{\partial z}{\partial f_r}\Bigr\rvert_{f = f_r},
\end{align}

where $\hat r$ and $\hat \theta$ are the directions corresponding to the dissipation and frequency quadratures, respectively \cite{Gao2008}. The projected data along the $\hat r$ direction was then used to calculate its corresponding PSD, $S_{\delta r}(\nu)$, according to Welch's method \cite{welch}. We used a Hann window with 50\% overlap in the calculation. The final conversion to a PSD in $\xi$ was found as

\begin{equation}
    S_\xi(\nu) = \frac{S_{\hat r}(\nu)}{|\partial z / \partial (1/Q_i)|^2_{f=f_r}} = \frac{Q_c^2}{Q^4|z_\infty|^2}S_{\hat r}(\nu).
\end{equation}

Note that we used three different frequency resolutions to plot $S_\xi(\nu)$ in Fig.~4a,b of the main text: \SI{0.1}{\Hz} for $\SI{0.1}{\Hz} \leq \nu < \SI{1}{\Hz}$, \SI{1}{\Hz} for $\SI{1}{\Hz} \leq \nu \leq \SI{10}{\Hz}$, and \SI{10}{\Hz} for $\SI{10}{\Hz} \leq \nu \leq \SI{250}{\Hz}$. To find the corresponding variance $\mathrm{Var}[\xi]$, we integrated $S_\xi(\nu)$ from \SI{0.1}{\Hz} to \SI{250}{\Hz} for each value of \nph. This corresponds to a variance over a time length of \SI{10}{\second}. 

\section{Fitting the empirical cumulative distribution functions}

The empirical cumulative distribution functions (eCDF) of $\xi$ at $\nph \sim 1$ in Fig.~5b,d of the main text were fitted to a log-normal distribution whose cumulative distribution function has the form

\begin{equation}
    f(\xi) = \frac{1}{2}\left[1 + \mathrm{erf}(\frac{\ln(\xi) - \mu}{\sigma\sqrt{2}}) \right],
    \label{eqn:lognormal}
\end{equation}

where erf is the error function, and $\mu$ and $\sigma$ are the mean and standard deviation of $\ln(\xi)$, respectively. Using the extracted $\mu$ and $\sigma$ parameters, we converted them to the corresponding mean and variance of the underlying $\xi$ as 

\begin{align}
    \mathrm{E}[\xi] &= e^{\mu + \sigma^2/2} \\
    \mathrm{Var}[\xi] &= (e^{\sigma^2}-1)e^{2\mu + \sigma^2}. 
\end{align}

We list the results of fitting the eCDFs to Eqn.~\ref{eqn:lognormal} in Table~\ref{table:ecdf} below.

\begin{table}[h]
    \centering
    \caption{Parameters obtained from fitting the eCDFs to a log-normal distribution}
    \begin{ruledtabular}
    \begin{tabular}{ccccc}
        %\toprule[1.5pt]
        N & $f_r$ (GHz) & $\mu[\ln(\xi/10^{-7})]$ & $\sigma[\ln(\xi/10^{-7})]$\\
        \specialrule{.5pt}{2pt}{2pt}
        1 & 5.43 & $1.311 \pm 0.004$ & $0.253 \pm .006$\\
        33 & 5.43 & $1.3899 \pm 0.0004$ & $0.0442 \pm .0006$ \\
        1 & 6.08 & $1.285 \pm 0.001$ & $0.186 \pm 0.002$\\
        37 & 6.09 & $1.341 \pm 0.0003$ & $0.0365 \pm .0005$
        \label{table:ecdf}
    \end{tabular}
    \end{ruledtabular}
\end{table}

\end{document}